\pdfoutput=1
\documentclass[conference]{IEEEtran}
\IEEEoverridecommandlockouts

\usepackage{times}

\usepackage[numbers]{natbib}
\usepackage{multicol}
\usepackage{hyperref}

\usepackage{amsmath}
\usepackage{amssymb}
\usepackage{amsfonts}
\usepackage{mathtools, nccmath}
\usepackage{xcolor}

\usepackage{siunitx}
\usepackage{cleveref}
\usepackage{graphicx}
\usepackage{booktabs}
\usepackage{soul}

\usepackage{amsthm}

\crefname{ass}{assumption}{assumptions}
\crefname{theorem}{theorem}{theorems}
\crefname{prop}{proposition}{propositions}
\crefname{lstlisting}{listing}{listings}
\Crefname{lstlisting}{Listing}{Listings}
\Crefname{figure}{Fig.}{Fig.}

\usepackage{listings}
\definecolor{dkgreen}{rgb}{0,0.6,0}
\definecolor{dred}{rgb}{0.545,0,0}
\definecolor{dblue}{rgb}{0,0,0.545}
\definecolor{lgrey}{rgb}{0.95,0.95,0.95}
\definecolor{yellow}{rgb}{1.0,1.0,0.0}
\definecolor{gray}{rgb}{0.4,0.4,0.4}
\definecolor{darkblue}{rgb}{0.0,0.0,0.6}
\definecolor{dpurple}{rgb}{0.5,0.0,0.5}
\lstdefinelanguage{cpp}{
    backgroundcolor=\color{lgrey},  
    basicstyle=\footnotesize \ttfamily \color{black},
    breakatwhitespace=false,       
    breaklines=true,               
    captionpos=b,                   
    commentstyle=\color{dkgreen},   
    deletekeywords={...},          
    escapeinside={(*@}{@*)},                  
    frame=single,                  
    language=C++,                
    keywordstyle=\color{purple},  
    morekeywords={BRIEFDescriptorConfig,string,TiXmlNode,DetectorDescriptorConfigContainer,istringstream,cerr,exit}, 
    identifierstyle=\color{black},
    stringstyle=\color{blue},      
    numbers=left,                 
    numbersep=5pt,                  
    numberstyle=\tiny\color{gray}, 
    rulecolor=\color{black},        
    showspaces=false,               
    showstringspaces=false,
    xleftmargin=1em,
    xrightmargin=0.4em,
    showtabs=false,
    frame=none,
    stepnumber=1,                   
    tabsize=2,                     
    title=\lstname,
    keywords=[1]{UNGAR_VARIABLE},
    keywords=[2]{UNGAR_LEAF_VARIABLE},
    keywords=[3]{UNGAR_BRANCH_VARIABLE},
    keywordstyle=[1]{\color{dpurple}},
    keywordstyle=[2]{\color{dpurple}},
    keywordstyle=[3]{\color{dpurple}},
    }
\definecolor{delectricblue}{RGB}{93, 117, 131}
\colorlet{lightdelectricblue}{delectricblue!30}
\colorlet{lightyellow}{yellow!30}

\newcommand{\hldb}[1]{%
    {%
    \sethlcolor{lightdelectricblue}%
    \hl{#1}%
    }%
}

\begin{document}

\title{\vspace*{18pt}Ungar -- A C++ Framework for Real-Time Optimal Control Using Template Metaprogramming}

\author{Flavio De Vincenti and Stelian Coros\thanks{This research was supported by the Swiss National Science Foundation through the National Centre of Competence in Digital Fabrication (NCCR DFAB) and through Grant No. 200021 200644, and by the European Research Council (ERC) under the European Union’s Horizon 2020 research and innovation program (Grant No. 866480).
        
We express our gratitude to Miguel Angel Zamora Mora and Zijun Hui for their precious contributions to the development of Ungar.

The authors are with the Computational Robotics Lab, ETH Zurich, Switzerland.
        {\tt\small flavio.devincenti@inf.ethz.ch}}}


%

\maketitle
\begin{abstract}
We present Ungar, an open-source library to aid the implementation of high-dimensional optimal control problems (OCPs). We adopt modern template metaprogramming techniques to enable the compile-time modeling of complex systems while retaining maximum runtime efficiency. Our framework provides syntactic sugar to allow for expressive formulations of a rich set of structured dynamical systems. While the core modules depend only on the header-only Eigen and Boost.Hana libraries, we bundle our codebase with optional packages and custom wrappers for automatic differentiation, code generation, and nonlinear programming. Finally, we demonstrate the versatility of Ungar in various model predictive control applications, namely, four-legged locomotion and collaborative loco-manipulation with multiple one-armed quadruped robots. Ungar is available under the Apache License 2.0 at \url{https://github.com/fdevinc/ungar}.
\end{abstract}

\IEEEpeerreviewmaketitle

\renewcommand{\S}{\mathbb{S}}
\newcommand{\R}{\mathbb{R}}
\renewcommand{\N}{\mathbb{N}}

\newcommand{\Iframe}{\mathcal{I}}
\newcommand{\Bframe}{\mathcal{B}}

\newcommand{\w}{\mathbf{w}}

\newcommand{\X}{\mathbf{X}}
\newcommand{\x}{\mathbf{x}}
\renewcommand{\xi}{\mathbf{x}_i}
\newcommand{\xk}{\mathbf{x}^k}
\newcommand{\xkpo}{\mathbf{x}^{k + 1}}
\newcommand{\xz}{\mathbf{x}_0}
\newcommand{\xo}{\mathbf{x}_1}
\newcommand{\xt}{\mathbf{x}_2}
\newcommand{\xN}{\mathbf{x}_N}
\newcommand{\xki}{\mathbf{x}_i^k}
\newcommand{\xkz}{\mathbf{x}_0^k}
\newcommand{\xko}{\mathbf{x}_1^k}
\newcommand{\xkt}{\mathbf{x}_2^k}
\newcommand{\xkR}{\mathbf{x}_R^k}

\newcommand{\U}{\mathbf{U}}
\renewcommand{\u}{\mathbf{u}}
\newcommand{\uk}{\mathbf{u}_k}
\newcommand{\uz}{\mathbf{u}_0}
\newcommand{\uo}{\mathbf{u}_1}
\newcommand{\ut}{\mathbf{u}_2}
\newcommand{\uNmo}{\mathbf{u}_{N-1}}
\newcommand{\uki}{\mathbf{u}_{k, i}}
\newcommand{\uko}{\mathbf{u}_{k, 1}}
\newcommand{\ukt}{\mathbf{u}_{k, 2}}
\newcommand{\ukR}{\mathbf{u}_{k, R}}
\newcommand{\ukif}{\mathbf{u}_{k, i, f}}
\newcommand{\ukilf}{\mathbf{u}_{k, i, \textsc{lf}}}
\newcommand{\ukirf}{\mathbf{u}_{k, i, \textsc{rf}}}
\newcommand{\ukilh}{\mathbf{u}_{k, i, \textsc{lh}}}
\newcommand{\ukirh}{\mathbf{u}_{k, i, \textsc{rh}}}
\newcommand{\ukih}{\mathbf{u}_{k, i, h}}
\newcommand{\fkif}{\mathbf{f}_{k, i, f}}
\newcommand{\rkif}{\mathbf{r}_{k, i, f}}
\newcommand{\fkih}{\mathbf{f}_{k, i, h}}
\renewcommand{\r}{\mathbf{r}}
\newcommand{\f}{\mathbf{f}}
\newcommand{\ta}{\boldsymbol{\tau}}
\renewcommand{\s}{\mathbf{s}}
\newcommand{\taukih}{\boldsymbol{\tau}_{k, i, h}}
\newcommand{\skif}{s_{k, i, f}}
\newcommand{\rkih}{\mathbf{r}_{k, i, h}}

\newcommand{\p}{\mathbf{p}}
\newcommand{\q}{\mathbf{q}}
\newcommand{\pdot}{\dot{\mathbf{p}}}
\newcommand{\Om}{\boldsymbol{\Omega}}
\newcommand{\pddot}{\ddot{\mathbf{p}}}
\newcommand{\Omdot}{\dot{\boldsymbol{\Omega}}}

\newcommand{\lf}{\textsc{lf}}
\newcommand{\rf}{\textsc{rf}}
\newcommand{\lh}{\textsc{lh}}
\newcommand{\rh}{\textsc{rh}}

\newcommand{\moi}{\mathbf{I}}

\newcommand{\Deltat}{\Delta t}
\newcommand{\kpo}{k + 1}

\newcommand{\ftexttt}[1]{\texttt{\frenchspacing#1}}

\section{Introduction}
\label{sec:introduction}

The advancements in model predictive control (MPC) methods have endowed robots with exceptional athletic skills. Recent displays of humanoid \cite{AtlasSpinJumpBD} and quadruped robots \cite{Bjelonic2022OfflineML, Grandia2022PerceptiveLT} have shown feats that were once prerogative to science fiction. However, significant engineering efforts are still necessary to make such practical MPC implementations possible. The solution of big nonlinear programming (NLP) problems at real-time rates clashes with the inherent high dimensionality and fast dynamics of mechanical systems. These conflicting aspects translate to onerous computational costs to be met in fractions of seconds, thus calling for complex data structures and ingenious software designs.

We seek to facilitate the manual endeavors flowing into the development of MPC controllers. In our vision, ease of use and relevance to a broad range of applications are of paramount importance. These objectives are only achievable by carefully averting any runtime computational overheads. At the same time, user interfaces must provide an intuitive syntax that mirrors standardized, mathematical formulations of optimal control problems (OCPs).

Achieving robust MPC performance is challenging in many ways. Efficient NLP solvers must be coupled with fast derivative computations. Given the large numbers and interrelationships of the state and control variables, manual implementation of first and second-order derivatives would result in a tedious, error-prone process. Also, although mature automatic differentiation (AD) and NLP libraries exist, most implementations require all the variables stacked in a single vector, which necessitates some index-keeping logic. This fact begs the question of what data structures could store them while guaranteeing zero-cost access operations and adaptability to different system designs.

With Ungar, we provide a metalanguage that addresses these modeling challenges. Our solution introduces constructs that significantly simplify the definition of the NLP problems typically arising in optimal control. We use template metaprogramming (TMP) to delegate the generation of the necessary implementations to the compiler, while users can focus exclusively on the architectural details of a desired MPC application. Our approach makes the transcription into code of structured variable sets seamless while encoding all hierarchical and indexing information at compile time. Consequently, all read/write operations acting on correspondingly created objects incur no additional costs, just like ad hoc programming solutions. Since the core of Ungar is header-only, its integration in C++ projects is effortless. However, we also include an optional interface to CppADCodeGen \cite{CppAD, CppADCodeGen} for automatically generating derivatives and an optional sequential quadratic programming (SQP) solver using OSQP \cite{osqp} as a backend; if enabled, all external dependencies are automatically downloaded through CMake. Finally, we illustrate the capabilities of Ungar by implementing MPC controllers for increasingly complex systems, including quadruped robots and teams of four-legged manipulators cooperatively carrying an object.

\begin{lstlisting}[float=*ht, language=cpp, label={lst:quadrotor}, caption={Decision variables of an OCP for controlling a quadrotor with Ungar.}, escapechar=|]
// Define integral constants.|\label{line:integral_consts_start}|
constexpr auto N          = 30_c;
constexpr auto NUM_ROTORS = 4_c;|\label{line:integral_consts_end}|

// Define "leaf" variables.|\label{line:leaf_start}|
constexpr auto position         = var_c<"position",         3>;
constexpr auto orientation      = var_c<"orientation",      Q>;
constexpr auto linear_velocity  = var_c<"linear_velocity",  3>;
constexpr auto angular_velocity = var_c<"angular_velocity", 3>;
constexpr auto rotor_speed      = var_c<"rotor_speed",      1>;|\label{line:leaf_end}|

// Define "branch" variables.|\label{line:branch_start}|
constexpr auto x = var_c<"x"> <<= (position, orientation, linear_velocity, angular_velocity);|\label{line:x}|
constexpr auto X = var_c<"X"> <<= (N + 1_c) * x;|\label{line:X}|
constexpr auto u = var_c<"u"> <<= NUM_ROTORS * rotor_speed;
constexpr auto U = var_c<"U"> <<= N * u;
constexpr auto decision_variables = var_c<"decision_variables"> <<= (X, U);|\label{line:branch_end}|
\end{lstlisting}

\subsection*{Related Work}

There exist many open-source packages to assist in the creation of MPC controllers. A noncomprehensive list includes frameworks for modeling, simulation, and optimization-based control of mechanical systems, such as the Control Toolbox \cite{Giftthaler2018TheCT} and its successor OCS2 \cite{OCS2}, Drake \cite{drake}, Crocoddyl \cite{mastalli20crocoddyl}, TOWR \cite{winkler18}, FROST \cite{Hereid2017FROST}, Quad-SDK \cite{abs:norby-quad-sdk-2022}, SymForce \cite{Martiros-RSS-22}, etc. These libraries are geared toward robotic applications and the OCPs they solve require specific structures. In contrast, NLP-oriented frameworks address more general classes of nonlinear optimization problems; notable examples are IPOPT \cite{Wchter2006OnTI}, ACADOS \cite{Houska2011ACADOTO, Verschueren2021}, PSOPT \cite{Becerra2010SolvingCO}, CasADi \cite{Andersson2019}, etc. Ungar complements the above libraries by providing novel system modeling features. In particular, it allows for quickly setting up the data structures required by widely adopted AD implementations and OCP solvers. Eventually, while MPC is its primary focus, the design of Ungar makes it suitable for any application requiring the solution of finite dimensional NLP problems.

\begin{figure}
    \centering
    \includegraphics[width=\linewidth]{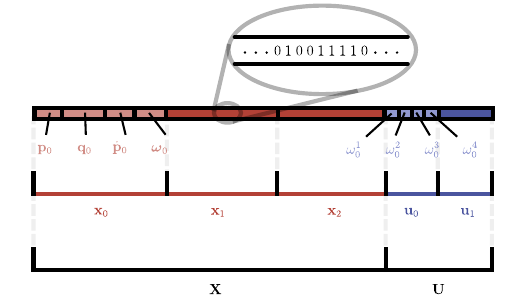}
    \caption{
    Hierarchical relationships among state and input variables of an OCP for controlling a quadrotor and underlying memory representation. At every time step $k$, we define the robot's state $\x_k$ as a stacked vector containing its position $\p_k \in \R^3$, orientation $\q_k \in \S^3 \subseteq \R^4$, linear velocity $\pdot_k \in \R^3$, and angular velocity $\boldsymbol{\omega}_k \in \R^3$. The inputs $\u_k$ consist of the four rotor speeds $\omega_k^i \in \R, \forall i \in \{1,2,3,4\}$. Finally, $\X \in \R^{13(N + 1)}$ and $\U^{4N}$ denote the stacked state and input vectors, where $N \in \N$ is the discrete time horizon---in the diagram, $N = 2$. Using TMP techniques, Ungar supports the generation of data structures for efficiently manipulating raw data arrays.
    }
    \label{fig:variables}
\end{figure}

\section{Data Structures}
\label{sec:ungar}

In this section, we introduce the two main data structures at the heart of Ungar: variables and maps. We accompany their descriptions with motivating examples in the robotics domain. We remark that the classes we discuss build on only two external dependencies, namely, Eigen \cite{eigenweb} and Boost.Hana \cite{hana}. The former is a linear algebra template library ubiquitous in robotics codebases due to its fast performance and versatility; the latter is a collection of utilities and containers that greatly simplify the implementation of TMP algorithms. Since both are header-only, we wrap them within Ungar to make its integration in C++ projects as straightforward as possible.

\subsection{Variables}
\label{subsec:variables}

At the core of Ungar lies the \verb|Variable| template class. Variables describe the structure of quantities of interest such as states, inputs, or parameters. Each variable has a name and a kind\footnote{In this paper, we assign distinct meanings to the terms \emph{kind} and \emph{type}. To clarify, \emph{kind} denotes the mathematical group to which a variable belongs, while \emph{type} exclusively refers to C++ data types.}, and it is related to all other variables through hierarchical relationships. When a \verb|Variable| object is instantiated, all this information is encoded in its type using three template parameters: a compile-time string for the name, an integer number identifying its kind, and a compile-time map representing the variable hierarchy. We employ \verb|boost::hana::string| and \verb|boost::hana::map| types to designate the name and the variable hierarchy, respectively. In particular, the latter maps variable names to sub-\verb|Variable| objects or arrays thereof.

To clarify and explain the above design choices, let us consider the decision variables of a finite horizon MPC for controlling a quadrotor. The state $\x_k$ of the system at time step $k$ consists of the robot's pose and velocity; the control inputs $\u_k$ comprise four rotor speeds. Given a discrete time horizon $N \in \mathbb{N}$ and assuming a direct multiple shooting formulation \cite{Diehl2005FastDM}, the optimization variables amount to the stacked state and input vectors $\X = [\x_0^\top \; \x_1^\top \; \ldots \x_N^\top]^\top$ and $\U = [\u_0^\top \; \u_1^\top \; \ldots \u_{N - 1}^\top]^\top$, respectively. Then, we can use the Ungar template object \verb|var_c| to instantiate the relevant variables as shown in \Cref{lst:quadrotor}.

The \verb|var_c| construct takes a fixed string denoting the name of the variable and an optional integer parameter representing its kind: if present, it instantiates a ``leaf'' variable, i.e., a variable that has no additional substructures; otherwise, it creates a ``branch'' variable, which is only defined in relation to its subvariables. We identify the kind of a leaf variable with \verb|1| for scalars, an implementation-defined constant \verb|Q| for unit quaternions, and any positive integer for a correspondingly sized vector. As schematically represented in \Cref{fig:variables}, Ungar allows mapping the structures encoded by variables to contiguous memory buffers (see \Cref{subsec:maps}).

\begin{lstlisting}[float=*ht, language=cpp, label={lst:bypass}, caption={Equivalent expressions for unambiguous variable hierarchies.}, escapechar=|]
static_assert(X(x, 1, linear_velocity)                     == X(linear_velocity, 1));
static_assert(U(u, 1, rotor_speed, 0)                      == U(rotor_speed, 1, 0));
static_assert(U(u, 1, rotor_speed, 1)                      == U(rotor_speed, 1, 1));
static_assert(decision_variables(X, x, 1, linear_velocity) == decision_variables(linear_velocity, 1));
static_assert(
    decision_variables(U, u, 2, rotor_speed, 3) == decision_variables(u, 2, rotor_speed, 3) &&
    decision_variables(U, u, 2, rotor_speed, 3) == decision_variables(rotor_speed, 2, 3)
);
\end{lstlisting}
\begin{lstlisting}[float=*ht, language=cpp, label={lst:quadrotor_macros}, caption={Equivalent transcription of \Cref{lst:quadrotor} with improved readability using macros.}, escapechar=|]
// Define integral constants.
constexpr auto N          = 30_c;|\label{line:integral_consts_start_macros}|
constexpr auto NUM_ROTORS = 4_c;|\label{line:integral_consts_end_macros}|

// Define "leaf" variables.
UNGAR_VARIABLE(position,           3);
UNGAR_VARIABLE(orientation,        Q);
UNGAR_VARIABLE(linear_velocity,    3);
UNGAR_VARIABLE(b_angular_velocity, 3);
UNGAR_VARIABLE(rotor_speed,        1);

// Define "branch" variables.
UNGAR_VARIABLE(x) <<= (position, orientation, linear_velocity, b_angular_velocity);
UNGAR_VARIABLE(X) <<= (N + 1_c) * x;
UNGAR_VARIABLE(u) <<= NUM_ROTORS * rotor_speed;
UNGAR_VARIABLE(U) <<= N * u;
UNGAR_VARIABLE(decision_variables) <<= (X, U);
\end{lstlisting}

In our example, the lines \ref{line:integral_consts_start}-\ref{line:integral_consts_end} define a discrete time horizon $N$ of $30$ time steps and the number of rotors as integral constants \cite{hana} through the user-defined literal \verb|_c|. The lines \ref{line:leaf_start}-\ref{line:leaf_end} define the leaf variables for the pose and velocity of a quadrotor, as well as its rotor speeds. Instead, the lines \ref{line:branch_start}-\ref{line:branch_end} introduce the branch variables for $\x_k$, $\X$, $\u_k$, and $\U$, respectively. We observe that $\x_k$ consists of the stacked position, orientation, linear velocity, and angular velocity of the robot, while $\X$ contains $N + 1$ stacked states. Then, we can immediately express the above structural information using the overloaded functions \verb|operator<<=|, \verb|operator,|, and \verb|operator*| as shown at lines \ref{line:x}-\ref{line:X}. Similarly, we define the branch variables for $\u_k$ and $\U$, and finally we stack $\X$ and $\U$ inside the object \verb|decision_variables| at line \ref{line:branch_end}.

All the variables defined in \Cref{lst:quadrotor} are \verb|constexpr| and can be queried at compile-time for two main pieces of information: their sizes and their indices within the hierarchy. For instance, with reference to the diagram in \Cref{fig:variables} and for $N = 30$, we can write
\begin{lstlisting}[language=cpp, captionpos=none]
static_assert(
    x.Size() == 13 &&
    X.Size() == 403 &&
    u.Size() == 4 &&
    U.Size() == 120 &&
    decision_variables.Size() == 523
);
\end{lstlisting}
and
\begin{lstlisting}[language=cpp, captionpos=none]
static_assert(
    X(x, 0).Index() == 0 &&
    X(x, 1).Index() == x.Size() &&
    X(x, 1, linear_velocity).Index() == 20 &&
    decision_variables(U).Index() == X.Size() &&
    U(u, 0).Index() == 0 &&
    U(u, 1).Index() == 4 &&
    U(u, 1, rotor_speed, 0).Index() == 4 &&
    U(u, 1, rotor_speed, 1).Index() == 5
);
\end{lstlisting}
As shown in the above listings, we can access all subvariables of any branch variable using the function call operator \verb|operator()|. If multiple copies of a subvariable exist, we must write the zero-based index of the subvariable we are interested in. Most importantly, if there is no ambiguity in the path from a branch variable to any of its subvariables, we can bypass all intermediate variables as shown in \Cref{lst:bypass}: this feature is very convenient when the variables are organized in deep hierarchies. Finally, Ungar provides macros for defining variables according to the convention that their name is identical to the corresponding object name. Thus, \Cref{lst:quadrotor} can be rewritten in a more succinct way as shown in \Cref{lst:quadrotor_macros}.

\subsection{Variable Maps}
\label{subsec:maps}

The \verb|Variable| framework provides the means to describe complex systems with a minimal yet expressive syntax without incurring runtime costs. To turn system descriptions into useful data structures, Ungar offers the template class \verb|VariableMap|. A variable map associates a variable with an array of scalars and each subvariable with a subarray. To perform the various mappings, we adopt the \verb|Eigen::Map| class \cite{eigenweb}, which allows interfacing raw buffers with dense matrix expressions seamlessly. All necessary Eigen maps are created during the execution of the \verb|VariableMap| constructor; therefore, accessing any subvariable data has no runtime cost. This is only possible due to our adoption of TMP techniques and makes Ungar akin to a metalanguage.

Given some user-defined scalar type \verb|scalar_t|, we can create a variable map for the quadrotor decision variables introduced in \Cref{subsec:variables} as:
\begin{lstlisting}[language=cpp, captionpos=none]
auto vars = MakeVariableMap<scalar_t>(decision_variables);
\end{lstlisting}
Then, we can access all submaps by passing corresponding subvariables to the \verb|Get| method. For example, we can initialize all unit quaternions to identity rotations and all remaining decision variables to zero \cite{eigenweb} as:
\begin{lstlisting}[language=cpp, captionpos=none, escapechar=|]
vars.Get(X).setZero();
for (auto k = 0; k < N + 1; ++k) {
    vars.Get(orientation, k).setIdentity();
}
vars.Get(U).setZero();
static_assert(|\label{line:assert_return_type_var_map}|
    std::same_as<
        decltype(vars.Get(U)),
        Eigen::Map<Eigen::VectorX<scalar_t>>&
    >
);
\end{lstlisting}
We remark that all objects returned by the \verb|Get| method have reference types, hence they do not perform copies and directly manipulate the underlying data. Also, the returned type depends on the kind of the corresponding variable, so it can be a reference to \verb|scalar_t|, an Eigen map to a unit quaternion, or an Eigen map to a vector. Branch variables are always mapped to vectors spanning all the corresponding subvariables (line \ref{line:assert_return_type_var_map}).

We enable this flexibility by internally adopting compile-time maps associating subvariables to corresponding data subarrays. While this solution ensures the best possible runtime performance, it can be demanding in terms of compile time and memory consumption. This can be undesirable if, for instance, such variable maps are employed only for intermediate code generation steps; indeed, most code generators optimize the input code, thus making any performance optimizations unnecessary at this stage. We address this need with a lazy version of \verb|VariableMap| named \verb|VariableLazyMap|. Lazy maps instantiate Eigen maps on demand, which is a cheap operation involving the copies of two integer numbers. Their constructors require a data buffer with the correct size as:
\begin{lstlisting}[language=cpp, captionpos=none]
VectorXr underlying{decision_variables.Size()};
auto lvars = MakeVariableLazyMap(underlying, decision_variables);
\end{lstlisting}
In particular, we can rewrite our initialization example as:
\begin{lstlisting}[language=cpp, captionpos=none, escapechar=|]
lvars.Get(X).setZero();
for (auto k = 0; k < N + 1; ++k) {
    lvars.Get(orientation, k).setIdentity();
}
lvars.Get(U).setZero();
static_assert(|\label{line:map_return_type}|
    std::same_as<
        decltype(lvars.Get(U)),
        Eigen::Map<Eigen::VectorX<scalar_t>>
    >
);
\end{lstlisting}
We highlight that the only difference between \verb|VariableMap| and \verb|VariableLazyMap| objects lies in the return types of \verb|Get|. As shown in the static assertion of line \ref{line:map_return_type}, lazy maps return Eigen maps by value instead of reference.
\section{Experiments}
\label{sec:experiments}

We validate Ungar by implementing two MPC schemes for different but related systems. The former is a quadrupedal locomotion controller based on the single rigid body dynamics (SRBD) model \cite{Bledt2019ImplementingRP}; the latter is a centralized MPC controller for collaborative locomotion and manipulation using one-armed quadruped robots \cite{DeVincenti2023CentralizedCLM}. In particular, the cooperative loco-manipulation example extends the simple locomotion controller to multiple arm-endowed robots carrying a shared payload. Our accompanying video presents recordings of simulation experiments using both controllers. These applications allow us to showcase the versatility of Ungar in defining completely different variable hierarchies by changing only a few lines of code. Although we do not provide the source code for the controllers we discuss, we bundle the library with multiple examples, including nonlinear MPC implementations for a quadrotor and a miniature radio-controlled car \cite{Liniger2015OptimizationbasedAR}.

\subsection{Implementation Details}
\label{subsec:implementation_details}

We implement our controllers relying exclusively on the functionalities provided by Ungar. For our tests, we employ the core data structures alongside the optional CppADCodeGen wrapper \cite{CppADCodeGen} and SQP solver. We base our solver on the recent work by \citet{Grandia2022PerceptiveLT} and refer the interested reader to \cite{DeVincenti2023CentralizedCLM} for more details.

To generate the necessary derivatives, CppADCodeGen requires all functions to depend on a single array of data including both independent variables and parameters. We comply with this interface through Ungar maps as described in \Cref{subsec:maps}. For each controller, we define the variable hierarchies \verb|decision_variables| and \verb|parameters|. The decision variables include states and control inputs, while the parameters contain inertial properties, reference trajectories, physical constants, etc. In the following sections, we will show the definitions of only the decision variables of the various MPC controllers to keep the exposition succinct. Nevertheless, we consistently adopt this subdivision in all our implementations, and we invite the reader to explore the examples in the library for a detailed walk-through.

Ungar requires compilers with C++20 support\footnote{For convenience, we provide a library version compliant with C++17 on a separate Git branch. Although this adaptation lacks some features, it implements all the most important functionalities of Ungar. We refer the reader to the documentation on the GitHub webpage for further information.}. We implemented and thoroughly tested our controllers on Ubuntu 22.04.2 LTS with GCC 11, but Ungar's core modules do not depend on any OS-specific instructions. However, we note that our optional CppADCodeGen wrapper uses runtime compilation and dynamic linking features offered by Linux.

\subsubsection*{Compile Times}

\begin{figure}[t]
    \centering
    \includegraphics[width=\linewidth]{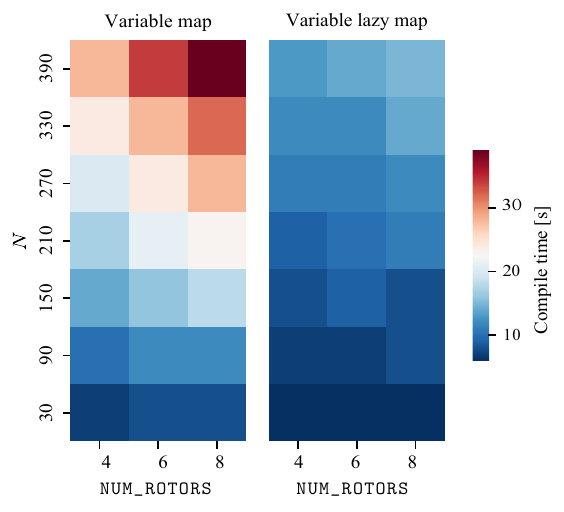}
    \caption{
    Compile times to generate the implementations of a \texttt{VariableMap} (\textbf{left}) and a \texttt{VariableLazyMap} (\textbf{right}) for the multirotor example of \Cref{lst:quadrotor_macros}. We benchmark Ungar against different time horizons and numbers of rotors by varying the lines \ref{line:integral_consts_start_macros} and \ref{line:integral_consts_end_macros}. The heatmaps manifest the more desirable compile times of the lazy map compared to its non-lazy alternative. For this reason, the \texttt{VariableMap} type should only be employed when seeking the best MPC performance possible.
    }
    \label{fig:heat_map}
\end{figure}

Ungar map objects have no runtime overheads by construction. Thus, we only provide a benchmark of the compile times for systems with different sizes. For this purpose, we measure the time required to build a \verb|VariableMap| and a \verb|VariableLazyMap| object for the decision variables defined in \Cref{lst:quadrotor_macros}. We take this measurement for different values of \verb|N| and \verb|NUM_ROTORS| on a laptop computer with an i7-11800H, \SI{2.30}{\giga\hertz}, 16-core CPU, and we plot corresponding heatmaps in \Cref{fig:heat_map}. We can see that the lazy map has more favorable compile times, requiring only \SI{14}{\second} to build the data structures for an octocopter with $N = 390$. In contrast, for the same setup, the compile time of \verb|VariableMap| is \SI{39}{\second} long. Although \verb|VariableMap| scales less favorably compared to \verb|VariableLazyMap|, we can notice that they have similar compile time performance for time horizons smaller than $100$ time steps. Nevertheless, we recommend using lazy maps for prototyping and switching to \verb|VariableMap| for production code to get the fastest runtime performance.

\subsection{Quadrupedal Locomotion}
\label{subsec:quadrupedal_locomotion}

We base our MPC formulation for quadrupedal locomotion on the controller of \citet{Bledt2019ImplementingRP}, but with three notable differences. Firstly, we use the nonlinear SRBD equations without linearizations or simplifications. Secondly, we represent orientations with unit quaternions instead of Euler angles to prevent singularity issues. Lastly, we employ a Lie group time-stepping method to integrate the dynamics conserving quaternion unit-norm constraints \cite{DeVincenti2023CentralizedCLM}.

We define our variable hierarchy as shown in \Cref{lst:loco}. In particular, we can see that less than $19$ lines of code suffice for generating the data structures required to manipulate the robot's states and inputs. While an MPC locomotion controller requires additional components to be of practical use, such as gait planners, inverse kinematics solvers, and whole-body controllers, we can already appreciate the potential of Ungar in simplifying the formulation of NLP problems with elaborate structures.

\subsection{Collaborative Loco-Manipulation}
\label{subsec:collaborative_loco_manipulation}

We use Ungar to implement an MPC controller that simultaneously optimizes ground reaction forces, manipulation wrenches, stepping locations, and body trajectories for a team of one-armed quadruped robots collectively manipulating an object. The resulting optimal control problem is very high-dimensional and presents coupled dynamics and deep hierarchies among states and inputs. For instance, each robot has $4$ legs, and each leg is associated with a ground reaction force and a stepping location at every time step; also, each robot has an arm that, in turn, corresponds to a manipulation force and torque.

We formulate our MPC for collaborative loco-manipulation (CLM) as an extension of the SRBD to multi-agent systems and refer the interested reader to \cite{DeVincenti2023CentralizedCLM} for a detailed description of our model. As shown in \Cref{lst:loco_manip}, the creation of Ungar variables for the seemingly involved CLM setting requires only minor changes to the locomotion control problem of \Cref{lst:loco}.

\subsection{Limitations}
\label{subsec:limitations}

The optimal runtime efficiency of Ungar is due to a large number of compile-time computations. However, if the variable hierarchies become too deep or nested, then compile times may become significantly high. Also, we observed compiler crashes when instantiating \verb|VariableMap| objects for very large OCPs, once again caused by the excessive amount of compile-time computations. In these rare cases, it is sufficient to adopt the \verb|VariableLazyMap| type, which is considerably less computation-intensive and provides almost the same performance as its non-lazy version. For future work, we will optimize the design of Ungar to improve its compile times. We additionally plan to expand the library with more tools to facilitate the implementation of high-performance MPC controllers.

\begin{lstlisting}[float=*ht, language=cpp, label={lst:loco}, caption={Decision variables of an OCP for quadrupedal locomotion using the single rigid body model.}]
// Define integral constants.
constexpr auto N        = 30_c;
constexpr auto NUM_LEGS = 4_c;

// Define "leaf" variables.
UNGAR_VARIABLE(position,          3);
UNGAR_VARIABLE(orientation,       Q);
UNGAR_VARIABLE(linear_velocity,   3);
UNGAR_VARIABLE(angular_velocity,  3);
UNGAR_VARIABLE(force,             3);
UNGAR_VARIABLE(relative_position, 3);

// Define "branch" variables.
UNGAR_VARIABLE(leg_input) <<= (force, relative_position);
UNGAR_VARIABLE(x)         <<= (position, orientation, linear_velocity, angular_velocity);
UNGAR_VARIABLE(X)         <<= (N + 1_c) * x;
UNGAR_VARIABLE(u)         <<= NUM_LEGS * leg_input;
UNGAR_VARIABLE(U)         <<= N * u;
UNGAR_VARIABLE(decision_variables) <<= (X, U);
\end{lstlisting}
\begin{lstlisting}[float=*ht, language=cpp, label={lst:loco_manip}, caption={Decision variables of an OCP for collaborative loco-manipulation with two robots modeled as single rigid bodies. We highlight the differen-\\ces from the locomotion controller formulated in \Cref{lst:loco}. In particular, we mark newly added variables in yellow and modified variables in light blue.}, escapechar=|]
// Define integral constants.
constexpr auto N          = 10_c;
constexpr auto |\hl{NUM\_ROBOTS}| = 2_c;|\label{line:num_robots}|
constexpr auto NUM_LEGS   = 4_c;

// Define "leaf" variables.
UNGAR_VARIABLE(position,          3);
UNGAR_VARIABLE(orientation,       Q);
UNGAR_VARIABLE(linear_velocity,   3);
UNGAR_VARIABLE(angular_velocity,  3);
UNGAR_VARIABLE(force,             3);
UNGAR_VARIABLE(relative_position, 3);
UNGAR_VARIABLE(|\hl{torque}|,            3);

// Define "branch" variables.
UNGAR_VARIABLE(leg_input)     <<= (force, relative_position);
UNGAR_VARIABLE(|\hl{arm\_input}|)     <<= (force, torque);
UNGAR_VARIABLE(|\hl{robot\_input}|)   <<= (NUM_LEGS * leg_input, arm_input);
UNGAR_VARIABLE(|\hl{payload\_state}|) <<= (position, orientation, linear_velocity, angular_velocity);
UNGAR_VARIABLE(|\hl{robot\_state}|)   <<= (position, orientation, linear_velocity, angular_velocity);
UNGAR_VARIABLE(|\hldb{x}|)             <<= (payload_state, NUM_ROBOTS * robot_state);
UNGAR_VARIABLE(X)             <<= (N + 1_c) * x;
UNGAR_VARIABLE(|\hldb{u}|)             <<= NUM_ROBOTS * robot_input;
UNGAR_VARIABLE(U)             <<= N * u;
UNGAR_VARIABLE(decision_variables) <<= (X, U);
\end{lstlisting}

\section{Conclusion}
\label{sec:conclusion}

In this paper, we introduced Ungar, a C++ template library for real-time MPC applications. Our framework uses TMP techniques to address modeling needs overlooked in existing NLP and optimal control software packages. In particular, it provides a metalanguage to describe complex systems in terms of variable hierarchies. Then, it lets compilers produce highly efficient code for manipulating raw data buffers based on these hierarchies. As shown in our quadruped locomotion and collaborative loco-manipulation experiments, these features enable great flexibility in formulating NLP problems and simplify AD-compliant implementations.



\bibliographystyle{plainnat}
\bibliography{references}

\begin{thebibliography}{25}
\providecommand{\natexlab}[1]{#1}
\providecommand{\url}[1]{\texttt{#1}}
\expandafter\ifx\csname urlstyle\endcsname\relax
  \providecommand{\doi}[1]{doi: #1}\else
  \providecommand{\doi}{doi: \begingroup \urlstyle{rm}\Url}\fi

\bibitem[Andersson et~al.(2019)Andersson, Gillis, Horn, Rawlings, and
  Diehl]{Andersson2019}
Joel A~E Andersson, Joris Gillis, Greg Horn, James~B Rawlings, and Moritz
  Diehl.
\newblock {CasADi} -- {A} software framework for nonlinear optimization and
  optimal control.
\newblock \emph{Mathematical Programming Computation}, 11\penalty0
  (1):\penalty0 1--36, 2019.
\newblock \doi{10.1007/s12532-018-0139-4}.

\bibitem[Becerra(2010)]{Becerra2010SolvingCO}
Victor~M. Becerra.
\newblock Solving complex optimal control problems at no cost with psopt.
\newblock \emph{2010 IEEE International Symposium on Computer-Aided Control
  System Design}, pages 1391--1396, 2010.

\bibitem[Bell()]{CppAD}
Bradley~M. Bell.
\newblock {CppAD}: a package for c++ algorithmic differentiation.
\newblock [Online]. Available: \url{https://github.com/coin-or/CppAD}.

\bibitem[Bjelonic et~al.(2022)Bjelonic, Grandia, Geilinger, Harley, Medeiros,
  Pajovic, Jelavic, Coros, and Hutter]{Bjelonic2022OfflineML}
Marko Bjelonic, Ruben Grandia, Moritz Geilinger, Oliver Harley, Vivian~Suzano
  Medeiros, Vuk Pajovic, Edo Jelavic, Stelian Coros, and Marco Hutter.
\newblock Offline motion libraries and online mpc for advanced mobility skills.
\newblock \emph{The International Journal of Robotics Research}, 41:\penalty0
  903 -- 924, 2022.

\bibitem[Bledt and Kim(2019)]{Bledt2019ImplementingRP}
G.~Bledt and Sangbae Kim.
\newblock Implementing regularized predictive control for simultaneous
  real-time footstep and ground reaction force optimization.
\newblock \emph{2019 IEEE/RSJ International Conference on Intelligent Robots
  and Systems (IROS)}, pages 6316--6323, 2019.

\bibitem[Diehl et~al.(2005)Diehl, Bock, Diedam, and Wieber]{Diehl2005FastDM}
M.~Diehl, H.~Bock, H.~Diedam, and Pierre-Brice Wieber.
\newblock Fast direct multiple shooting algorithms for optimal robot control.
\newblock 2005.

\bibitem[Dionne et~al.()]{hana}
Louis Dionne et~al.
\newblock {Boost.Hana}: {Y}our standard library for metaprogramming.
\newblock [Online]. Available: \url{https://github.com/boostorg/hana}.

\bibitem[Dynamics(2023)]{AtlasSpinJumpBD}
Boston Dynamics.
\newblock {A}tlas {G}ets a {G}rip | {B}oston {D}ynamics.
\newblock \url{https://www.youtube.com/watch?v=-e1_QhJ1EhQ}, January 2023.

\bibitem[Farshidian et~al.()]{OCS2}
Farbod Farshidian et~al.
\newblock {OCS2}: An open source library for optimal control of switched
  systems.
\newblock [Online]. Available: \url{https://github.com/leggedrobotics/ocs2}.

\bibitem[Giftthaler et~al.(2018)Giftthaler, Neunert, St{\"a}uble, and
  Buchli]{Giftthaler2018TheCT}
Markus Giftthaler, Michael Neunert, M.~St{\"a}uble, and Jonas Buchli.
\newblock The control toolbox — an open-source c++ library for robotics,
  optimal and model predictive control.
\newblock \emph{2018 IEEE International Conference on Simulation, Modeling, and
  Programming for Autonomous Robots (SIMPAR)}, pages 123--129, 2018.

\bibitem[Grandia et~al.(2022)Grandia, Jenelten, Yang, Farshidian, and
  Hutter]{Grandia2022PerceptiveLT}
Ruben Grandia, Fabian Jenelten, Shao-Hua Yang, Farbod Farshidian, and Marco
  Hutter.
\newblock Perceptive locomotion through nonlinear model predictive control.
\newblock \emph{ArXiv}, abs/2208.08373, 2022.

\bibitem[Guennebaud et~al.(2010)Guennebaud, Jacob, et~al.]{eigenweb}
Ga\"{e}l Guennebaud, Beno\^{i}t Jacob, et~al.
\newblock Eigen v3.
\newblock http://eigen.tuxfamily.org, 2010.

\bibitem[Hereid and Ames(2017)]{Hereid2017FROST}
Ayonga Hereid and Aaron~D. Ames.
\newblock Frost: Fast robot optimization and simulation toolkit.
\newblock In \emph{IEEE/RSJ International Conference on Intelligent Robots and
  Systems (IROS)}, Vancouver, BC, Canada, September 2017. IEEE/RSJ.

\bibitem[Houska et~al.(2011)Houska, Ferreau, and Diehl]{Houska2011ACADOTO}
Boris Houska, Hans~Joachim Ferreau, and Moritz Diehl.
\newblock Acado toolkit—an open‐source framework for automatic control and
  dynamic optimization.
\newblock \emph{Optimal Control Applications and Methods}, 32, 2011.

\bibitem[Leal et~al.()]{CppADCodeGen}
Joao~Rui Leal et~al.
\newblock {CppADCodeGen}.
\newblock [Online]. Available: \url{https://github.com/joaoleal/CppADCodeGen}.

\bibitem[Liniger et~al.(2015)Liniger, Domahidi, and
  Morari]{Liniger2015OptimizationbasedAR}
Alexander Liniger, Alexander Domahidi, and Manfred Morari.
\newblock Optimization‐based autonomous racing of 1:43 scale rc cars.
\newblock \emph{Optimal Control Applications and Methods}, 36:\penalty0 628 --
  647, 2015.
\newblock URL \url{https://api.semanticscholar.org/CorpusID:11242645}.

\bibitem[Martiros et~al.(2022)Martiros, Miller, Bucki, Solliday, Kennedy, Zhu,
  Dang, Pattison, Zheng, Tomic, Henry, Cross, VanderMey, Sun, Wang, and
  Holtz]{Martiros-RSS-22}
Hayk Martiros, Aaron Miller, Nathan Bucki, Bradley Solliday, Ryan Kennedy, Jack
  Zhu, Tung Dang, Dominic Pattison, Harrison Zheng, Teo Tomic, Peter Henry,
  Gareth Cross, Josiah VanderMey, Alvin Sun, Samuel Wang, and Kristen Holtz.
\newblock {SymForce: Symbolic Computation and Code Generation for Robotics}.
\newblock In \emph{Proceedings of Robotics: Science and Systems}, 2022.
\newblock \doi{10.15607/RSS.2022.XVIII.041}.

\bibitem[Mastalli et~al.(2020)Mastalli, Budhiraja, Merkt, Saurel, Hammoud,
  Naveau, Carpentier, Righetti, Vijayakumar, and Mansard]{mastalli20crocoddyl}
Carlos Mastalli, Rohan Budhiraja, Wolfgang Merkt, Guilhem Saurel, Bilal
  Hammoud, Maximilien Naveau, Justin Carpentier, Ludovic Righetti, Sethu
  Vijayakumar, and Nicolas Mansard.
\newblock {Crocoddyl: An Efficient and Versatile Framework for Multi-Contact
  Optimal Control}.
\newblock In \emph{IEEE International Conference on Robotics and Automation
  (ICRA)}, 2020.

\bibitem[Norby et~al.(2022)Norby, Yang, Tajbakhsh, Ren, Yim, Stutt, Yu,
  Flowers, and Johnson]{abs:norby-quad-sdk-2022}
Joseph Norby, Yanhao Yang, Ardalan Tajbakhsh, Jiming Ren, Justin~K. Yim,
  Alexandra Stutt, Qishun Yu, Nikolai Flowers, and Aaron~M. Johnson.
\newblock Quad-{SDK}: Full stack software framework for agile quadrupedal
  locomotion.
\newblock In \emph{ICRA Workshop on Legged Robots}, May 2022.

\bibitem[Stellato et~al.(2020)Stellato, Banjac, Goulart, Bemporad, and
  Boyd]{osqp}
B.~Stellato, G.~Banjac, P.~Goulart, A.~Bemporad, and S.~Boyd.
\newblock {OSQP}: an operator splitting solver for quadratic programs.
\newblock \emph{Mathematical Programming Computation}, 12\penalty0
  (4):\penalty0 637--672, 2020.
\newblock \doi{10.1007/s12532-020-00179-2}.
\newblock URL \url{https://doi.org/10.1007/s12532-020-00179-2}.

\bibitem[Tedrake and the Drake Development~Team(2016)]{drake}
Russ Tedrake and the Drake Development~Team.
\newblock Drake: A planning, control, and analysis toolbox for nonlinear
  dynamical systems, 2016.
\newblock URL \url{http://drake.mit.edu}.

\bibitem[Verschueren et~al.(2021)Verschueren, Frison, Kouzoupis, Frey, van
  Duijkeren, Zanelli, Novoselnik, Albin, Quirynen, and Diehl]{Verschueren2021}
Robin Verschueren, Gianluca Frison, Dimitris Kouzoupis, Jonathan Frey, Niels
  van Duijkeren, Andrea Zanelli, Branimir Novoselnik, Thivaharan Albin, Rien
  Quirynen, and Moritz Diehl.
\newblock acados -- a modular open-source framework for fast embedded optimal
  control.
\newblock \emph{Mathematical Programming Computation}, Oct 2021.
\newblock ISSN 1867-2957.
\newblock \doi{10.1007/s12532-021-00208-8}.
\newblock URL \url{https://doi.org/10.1007/s12532-021-00208-8}.

\bibitem[Vincenti and Coros(2023)]{DeVincenti2023CentralizedCLM}
Flavio~De Vincenti and Stelian Coros.
\newblock Centralized model predictive control for collaborative
  loco-manipulation.
\newblock In \emph{Robotics: Science and Systems}, 2023.
\newblock URL \url{https://api.semanticscholar.org/CorpusID:259319056}.

\bibitem[W{\"a}chter and Biegler(2006)]{Wchter2006OnTI}
Andreas W{\"a}chter and Lorenz~T. Biegler.
\newblock On the implementation of an interior-point filter line-search
  algorithm for large-scale nonlinear programming.
\newblock \emph{Mathematical Programming}, 106:\penalty0 25--57, 2006.

\bibitem[Winkler et~al.(2018)Winkler, Bellicoso, Hutter, and Buchli]{winkler18}
Alexander~W Winkler, Dario~C Bellicoso, Marco Hutter, and Jonas Buchli.
\newblock Gait and trajectory optimization for legged systems through
  phase-based end-effector parameterization.
\newblock \emph{IEEE Robotics and Automation Letters (RA-L)}, 3:\penalty0
  1560--1567, July 2018.
\newblock \doi{10.1109/LRA.2018.2798285}.

\end{thebibliography}

\end{document}